\documentclass{ws-ijmpa}

\begin{document}

\markboth{I.Milillo, M.Lattanzi, G.Montani} {On the coupling
between Spin and Cosmological Gravitational Waves }

\catchline{}{}{}{}{}
%
%%%%%%%%%%%%%%%%%%%%%%%%%%%%%%%%%%%%%%%%%%%%%%%%%%%%%%%%%%%%%%%%%%%%

\title{ON THE COUPLING BETWEEN SPINNING PARTICLES AND COSMOLOGICAL GRAVITATIONAL WAVES 
}

\author{IRENE MILILLO}

\address{Institute of Cosmology and Gravitation, University of Portsmouth\\Mercantile House, Hampshire Terrace
Portsmouth, Hants\\ PO1 2EG, UK \\
Irene.Milillo@port.ac.uk}

\author{MASSIMILIANO LATTANZI}

\address{Oxford Astrophysics\\
Denys Wilkinson Building, Keble Road,
OX3 8ND, Oxford, UK\\and\\
Istituto Nazionale di Fisica Nucleare \\ 
Via Enrico Fermi , 40  - 00044 
Frascati (Rome) Italy \\
mxl@astro.ox.ac.uk}

\author{GIOVANNI MONTANI}

\address{Dipartimento di Fisica and ICRA, Sapienza, University of Rome,\\Piazzale Aldo Moro, 5- 00185 Roma, Italy\\
ENEA -- C.R. Frascati (Department F.P.N.),\\Via 
Enrico Fermi, 45- 00044, Frascati (Roma), Italy\\
ICRANet -- C. C. Pescara,\\Piazzale della 
Repubblica, 10- 65100, Pescara, Italy\\
montani@icra.it}

\maketitle

%\begin{history}
%\received{Day Month Year}
%\revised{Day Month Year}
%\end{history}

\begin{abstract}
The influence of spin in a system of classical particles on the propagation
of gravitational waves is
analyzed in the cosmological context of primordial thermal
equilibrium. On a flat Friedmann-Robertson-Walker metric, when the precession is
neglected, there is no contribution due to the spin to the
distribution function of the particles. Adding a small
tensor perturbation to the background metric, we study if a coupling
between gravitational waves and spin exists that can modify the evolution of
the distribution function, leading to new terms in the anisotropic
stress, and then to a new source for gravitational waves. In the chosen
gauge, the final result is that, in the absence of other kind of perturbations,
 there is no coupling between spin and gravitational waves.

\end{abstract}

%\ccode{PACS numbers: 11.25.Hf, 123.1K}

\section{Papapetrou Equations in the Isotropic and Homogenous Flat Universe}

We study the influence of spinning particles on propagation of  gravitational waves
over a cosmological background.
The equations of motion for a spinning particle in the framework of general relativity were derived by Papapetrou in 1951. He used a multipole expansion for the energy-momentum tensor of the particle such that, at dipole order, a deviation from geodesic motion and an equation describing spin precession are obtained.\cite{papo}\cdash\cite{papt} These equations are\\
\begin{eqnarray}
\frac{D}{Ds}p^{\mu}&=&-\frac{1}{2}R^{\mu}_{\;\nu\rho\sigma}S^{\rho\sigma}u^{\nu}\quad\label{1},
\label{eq:papa1}\\
\frac{D}{Ds}S^{\mu\nu}&=&p^{\mu}u^{\nu}-p^{\nu}u^{\mu},
\label{eq:papa2}
\end{eqnarray}\\
where $ds$ is the affine parameter, the vector $p^{\mu}$ is the
generalized momentum, the antisymmetric tensor $S^{\mu\nu}$ is the
angular momentum (spin) and $u^{\mu}=dx^{\mu}/ds$. In order to close the system we need to impose a supplementary condition
that determines the center of mass of the spinning particle; we
choose the Papapetrou condition:
\begin{equation}
S^{i0}=0.
\end{equation}
Other common choices that can be found in the literature are the Pirani condition ($S^{\mu \nu}u_{\nu}=0$) and the Tulkzyjew condition ($S^{\mu \nu}p_{\nu}=0$), the former coinciding with the Papapetrou condition in the rest frame of the particle.\\
In our work we consider a flat Friedmann-Robertson-Walker (FRW) background described by the
metric
\begin{equation}
 ds^{2}=-dt^{2}+a^{2}(t)(dx^{2}+dy^{2}+dz^{2}),
\label{eq:FRW} 
\end{equation}
 and assume that precession is absent \cite{khri}, so that Eq. (\ref{eq:papa2}) becomes:
\begin{equation}
\frac{D}{Ds}S^{\mu\nu}=0,
\end{equation}
and the generalized momentum coincides with the standard one.\\
Solving the Papapetrou equations in this case we obtain the
time dependence of the angular momentum tensor, in terms of the
scale factor:
\begin{equation}
S^{ij}=\frac{1}{a^{2}}\Sigma^{ij},
\label{eq:ang-mom}
\end{equation}
where $\Sigma^{ij}$ does not depend on time. For what concerns the momentum 4-vector,
the equation of motion results
being the standard geodesic equation, due to the symmetries of the
homogeneous and isotropic FRW metric.

\section{Boltzmann Equation: Influence of Spin-metric Terms}
The Boltzmann equation for a non-collisional system of particles, when the spin is included through Eq. (\ref{eq:papa1}), is
\begin{equation}
\frac{\partial{f}}{\partial{t}}+\frac{\partial{f}}{\partial{x^i}}\frac{p^i}{p^0}+\frac{\partial{f}}{\partial{p_i}}\frac{p^j
p^k}{p^0}\frac{\partial
g_{jk}}{\partial{x^i}}-\frac{1}{2p^0}\frac{\partial f}{\partial
p_i}(\frac{1}{2}R_{i \alpha \beta \gamma}p^{\alpha}S^{\beta
\gamma})=0,
\end{equation}\\
where $ f(x^i,p_j,t)$ is the distribution function of the fluid and $p^i$ and $p^0$ are expressed in terms of the independent
variable $p_i$ through the metric tensor.
Defining $p=\sqrt{p_ip_i}$ so that ${\partial p}/{\partial
p_i}=\hat{p}_i$ (hats denotes unit vectors) we can see that in the case of the FRW metric (\ref{eq:FRW}), the
coupling term is equal to zero, because of the vanishing of $S^{i0}$ and of the
antisymmetry of Riemann tensor:
\begin{equation}
-\frac{1}{2p^0}\frac{\partial f}{\partial p^i}R_{i\alpha \beta
\gamma}p^{\alpha}S^{\beta \gamma}=-\frac{1}{2p^0}\frac{\partial
f}{\partial p}(R_{i0 \beta \gamma}p^{0}\hat{p}_iS^{\beta
\gamma}+R_{ij \beta \gamma}p^{j}\hat{p}_iS^{\beta \gamma}).
\end{equation}
Then we conclude that on an unperturbed FRW background the spin does not influence the evolution of the distribution function.\\

\section{Spin-Gravitational Waves Coupling}
The next step is to consider small tensor perturbations of the metric and see if in this case the presence of the spin can influence the time dependence of the distribution function, through the analysis of the perturbed Boltzmann equation. The solution will give the first order variation of the distribution function from which we can compute the anisotropic stress tensor acting as source in the differential equations for the tensor metric perturbation.\\
Let us consider a perturbation of the spatial metric:
\begin{equation}
g_{ij}= a^{2}[\delta_{ij}+h_{ij}(\vec{x},t)],
\end{equation}
while the other components are left unperturbed: $g_{00}=-1$ and
$g_{0i}=0$. This amounts to chosing the synchronous gauge to study the 
evolution of perturbations. Here $h_{ij}(\vec{x},t) $ is a small perturbation that
satisfies the following conditions:
\begin{equation}
h_{ii}=0,\qquad
\partial_ih_{ij}=0.\qquad
\end{equation}
i.e. it is traceless and divergenceless and thus represents a gravitational wave propagating over
the FRW background.
Considering the variations of the metric, the distribution function and the spin tensor, in the Fourier space the Boltzmann equation at first order is
\begin{equation}
\dot{\delta{f}}+i\hat{p_{i}}\hat{k_{i}}\delta{f}+\frac{ip}{2}\frac{\partial{f^{(0)}}}{\partial{p}}\hat{p_{i}}\hat{p_{j}}\hat{p_{k}}\hat{k_{i}}(h_{jk}(u)-h_{jk}(0))-\frac{1}{2}\frac{\partial{f^{(0)}}}{\partial{p}}\frac{a}{k}\delta
R_{i0lk}\hat{p_{i}}S^{lk}=0,
\label{eq:boltz-first}
\end{equation}
where dot denotes derivative with respect to the time variable $u \equiv k\int_{t_1}^t dt'/a(t')$ and hats denote unit vectors; $f^{(0)}$ is the zeroth-order part of $f(\vec{x},\vec{p},t_1)$, which we assumed to have the
usual ideal gas form.
In computing this formula we have used the previous results about the unperturbed quantities. \\
The variation of the Riemann tensor in synchronous gauge is
\begin{equation}
\delta {R_{i0lk}}=\frac{ka}{2}(\dot{h}_{il,k}-\dot{h}_{ik,l})
\end{equation}
so that using Eq. (\ref{eq:ang-mom}) we can write Eq. (\ref{eq:boltz-first}) in the form
\begin{equation}
\dot{\delta{f}}+i\hat{p_{i}}\hat{k_{i}}\delta{f}+\frac{ip}{2}\frac{\partial{f^{(0)}}}{\partial{p}}\hat{p_{i}}\hat{p_{j}}\hat{p_{k}}\hat{k_{i}}(h_{jk}(u)-h_{jk}(0))-\frac{i}{2}\hat{p_{i}}\frac{\partial{f^{(0)}}}{\partial{p}}\Sigma^{lj}k_{j}\dot{h}_{il}=0.
\end{equation}
This is a first order differential equation that enables us to write a formal solution for $\delta f$ through integration with respect to $u$. We use this solution to compute the anisotropic stress part $\pi_{ij}$ of the energy-momentum tensor
$T^i_j$.
\begin{eqnarray}
\pi_{ij}&=&T^{i}_{j}-\overline{p}\delta_{ij}\\
T^{i}_{j}&=&\frac{1}{\sqrt{-g}}\int
f(x^i,p_j,t)\frac{p^ip_j}{p^0}d^3p
\end{eqnarray}
with $\overline{p}$ being the unperturbed pressure.\\
The final result is the sum of two different contributions
\begin{equation}
\pi_{ij}(u)=\pi_{ij}^{(0)}(u)+\pi_{ij}^{(S)}(u),
\end{equation}
the first term is spin-independent\cite{wein}\cdash\cite{latt} and the second is the one we are interested in. We find 
\begin{equation}
\pi^{(S)}_{ij}=\frac{i}{2}n\int^u_0du'K(u-u')k_m\Sigma^{lm}(k_i\dot
h_{jl}+k_j\dot h_{il})
\end{equation}
where the integral kernel
$K(u-u')(s)=\frac{1}{64}\int^1_{-1}e^{ixs}(1-x^2)x^2dx$ 
and $n$ is the particle number density. If we choose the $z$ axis to coincide
with the direction of propagation of the gravitational wave, 
we have that $\vec{k}=(0,0,1)$ and consequently, from the divergenceless of $h_{ij}$,
that $h_{i3}=0$. Then it is easy to see that
 the only components of $\pi^{(S)}_{ij}$ that are different from zero are those with one
index equal to~3.\\
The evolution equation for gravitational waves is
\begin{equation}
\ddot h_{ij}+2\frac{\dot a}{a}\dot h_{ij}+h_{ij}=16\pi
G\pi_{ij}.
\end{equation}
We can immediately see that, if only tensor modes (for which $h_{i3}=0$) are present, the non-vanishing components of the spin dependent part of the anisotropic stress do not enter in the equation. Then there is no coupling between the 
gravitational waves and the spin of the particles in the fluid.\\
The next step in this work will be to study of supplementary conditions other that the Papapetrou condition, and the inclusion of scalar and tensor modes in the analysis. The existence of a coupling could be interesting in the perspective ot the detection of cosmological gravitational waves by interferometers.

%\begin{thebibliography}{000} %for 3 digits
%\begin{thebibliography}{00}  %for 2 digits


\begin{thebibliography}{0}    %for 1 digit

%%journal paper   
\bibitem{papo} A. Papapetrou, {\it Proc. R. Soc. Lond.} {\bf A 209}, 248 (1951).
 

\bibitem{papt} E. Corinaldesi, A. Papapetrou, {\it Proc. R. Soc. Lond.} {\bf A 209 }, 259 (1951).


\bibitem{khri} I.B.Khriplovich, A.A.Pomerasky, {\it Survey High Energ.Phys} {\bf 14}, 145173 (1999).


\bibitem{wein} S.Weinberg, {\it Phys.Rev.D} {\bf 69}, 023503(2004).


\bibitem{latt} M.Lattanzi, G.Montani. {\it Mod.Phys.Lett. A} {\bf 20}, 2607 (2005).

\end{thebibliography}
\end{document}